\documentstyle[aps,prl,epsf,floats]{revtex}
\begin{document}
\draft
\twocolumn[\hsize\textwidth\columnwidth\hsize\csname@twocolumnfalse%
\endcsname

\title{A model for anomalous directed percolation}

\author{Haye Hinrichsen$^1$ and Martin Howard$^2$}

\address{$^1$ Max-Planck-Institut f\"ur Physik komplexer Systeme,
         N\"othnitzer Stra\ss e 38, D-01187 Dresden, Germany \\ 
         $^2$ CATS, The Niels Bohr Institute, Blegdamsvej 17,
         2100 Copenhagen \O, Denmark \\}

\date{\today}
\maketitle

\begin{abstract}
We introduce a model for the spreading of epidemics by long-range 
infections and investigate the critical behaviour at the
spreading transition. The model generalizes directed bond
percolation and is characterized by a probability 
distribution for long-range infections which 
decays in $d$ spatial dimensions as $1/r^{d+\sigma}$. 
Extensive numerical simulations are performed in 
order to determine the density exponent $\beta$
and the correlation length exponents $\nu_{||}$ and $\nu_\perp$
for various values of $\sigma$. We observe that 
these exponents vary continuously with $\sigma$, in
agreement with recent field-theoretic predictions.
We also study a model for pairwise annihilation of
particles with algebraically distributed long-range 
interactions.
\end{abstract}

] 

%
%
%


\parskip 2mm

\section{Introduction}
\label{Introduction}

Spreading processes are often encountered in nature in situations
as diverse as epidemics~\cite{mollis,liggett}, 
catalytic reactions~\cite{cataly},
forest fires~\cite{fires}, and transport in random 
media~\cite{rmedia1,rmedia2}. Depending on the particular environmental
conditions, the spreading process may either continue to
spread over the whole population or die out after some 
time. The essential features of this transition 
between survival and extinction of the spreading agent
may be described by simple stochastic lattice models, 
which mimic the spreading mechanism by certain probabilistic rules. 
Usually such models incorporate two competing processes, 
namely spreading (infection) of nearest neighbors and 
spontaneous recovery (healing), with or without immunization.
The spreading properties depend on the 
relative rates of the two processes. For example,
if the rate for infection is very low, the spreading agent
will disappear after some time and the system becomes trapped
in an inactive state (or set of states) which is usually
referred to as the {\it absorbing} state of the model. 
On the other hand, if infections occur more frequently, 
the spreading process may survive for a very long time. 
The main theoretical interest in these models
stems from the fact that the phase transition 
from the fluctuating active phase into the
non-fluctuating absorbing state is 
continuous, and characterized by
universal scaling laws associated with 
certain critical exponents. 
As in equilibrium  statistical mechanics, 
these exponents allow one to categorize different lattice 
models into universality classes. Each of these 
universality classes then corresponds to a specific underlying 
field theory.

The most important universality class for spreading 
transitions with short-range interactions is
Directed Percolation (DP)~\cite{dpreview}, as described
by Reggeon field theory~\cite{regfth1,regfth2,regfth3}. 
In fact, DP covers the majority of phase transitions 
from a fluctuating active phase into 
absorbing states. The fundamental properties of DP have
been expressed as a conjecture in Refs.~\cite{regfth3,dpconj}. 
Accordingly a spreading transition belongs to the DP class if 
(a)~the absorbing state is unique, 
(b)~the active phase is characterized by a one-component
positive order parameter, 
(c)~there are no other symmetries of the physical system except for
spatio-temporal translation and spatial reflection invariance, 
(d)~there is no frozen disorder, and
(e)~the dynamical rules involve only short-range interactions. 

In many realistic spreading processes, however, short-range 
interactions do not appropriately describe the 
underlying transport mechanism. This situation emerges, for example, 
when an infectious disease is transported by insects. The motion
of the insects is typically not a random walk, rather one observes
occasional flights over long distances before the next infection
occurs. Similar phenomena are expected when the spreading 
agent is subjected to a turbulent flow. It is intuitively 
clear that occasional spreading over long distances will 
significantly alter the spreading properties.
On a theoretical level such a {\it super-diffusive} 
motion may be described by L\'evy flights~\cite{rmedia2}, 
i.e., by uncorrelated random moves over 
algebraically distributed  distances. 

Anomalous directed percolation, as originally
proposed by Mollison~\cite{mollis}
in the context of epidemic spreading,
is a generalization of DP in which the spreading agent 
performs L\'evy flights. This means that the distribution
of the spreading distance $r$ is given by
\begin{equation}
\label{ProbDis}
P(r) \sim 1/r^{d+\sigma} \ ,
\qquad (\sigma>0) \ ,
\end{equation}
where $d$ denotes the spatial dimension of the system.
The exponent $\sigma$ is a free parameter that
controls the characteristic shape of the distribution.
It should be emphasized that $\sigma$ does {\it not} introduce
any new length scale, rather it changes the scaling properties 
of the underlying (anomalous) diffusion process. 
We are particularly interested in the critical 
properties of anomalous DP close 
to the phase transition. As in the case of ordinary
DP, we expect anomalous DP to be characterized 
by the universal critical exponents
$\beta$, $\nu_{\perp}$, and $\nu_{||}$. 
The exponent $\beta$ is related to the 
order parameter, the density of active sites $n$.
Since the DP process is controlled by a single parameter $p$,
with the phase transition taking place at $p=p_c$, then close to this 
transition in the active phase $n$ vanishes
as $n \sim (p-p_c)^\beta$. At the same time, 
we expect the spatial and temporal correlation lengths
$\xi_\perp$ and $\xi_{||}$ to diverge as 
$\xi_\perp \sim |p-p_c|^{-\nu_\perp}$ and 
$\xi_{||} \sim |p-p_c|^{-\nu_{||}}$, respectively, on {\it both} sides
of the transition. Theoretically, one is interested in the 
dependence of these exponents on $\sigma$, whether the exponents
are independent from one
another, and how they cross over to the exponents of ordinary DP.
Some time ago Grassberger~\cite{grassb} claimed that the critical
exponents of anomalous DP should depend continuously on the control
exponent $\sigma$. Very recently this work has been considerably
clarified and extended by Janssen~et.~al.~\cite{janssn}, who have
presented a comprehensive field-theoretic renormalization 
group (RG) calculation for anomalous spreading processes 
with and without immunization. The aim of the present work
is to verify their results numerically.
To this end we introduce a model for anomalous DP
which generalizes directed bond percolation. In contrast to
previously studied models~\cite{albano,cannas} we do not
introduce an upper cutoff for the flight distance $r$,
and hence finite size effects are drastically reduced.
Extensive numerical simulations are performed 
in order to determine the critical exponents,
which are found to compare favourably with the 
field-theoretic predictions.

The paper is organized as follows. In Section~\ref{FieldTheory}
we first review the known field-theoretic results.
In Section~\ref{LatticeModel} we introduce a lattice model 
for anomalous DP and discuss the role of finite size effects.
In Section~\ref{Numerics} the numerical results are presented and 
compared with the field-theoretic predictions.
We also discuss the case of anomalous pair annihilation 
in Section~\ref{Annihilation}.

\section{Field-theoretic predictions}
\label{FieldTheory}

In this section we will summarize some of the field-theoretic results 
which have been derived in Ref.~\cite{janssn}.
First of all let us recall that the Langevin equation 
for ordinary DP~\cite{regfth3} is given by
\begin{equation}
\frac{\partial}{\partial t} n({\bf x},t) =
(\tau+ D_N \nabla^2) \, n({\bf x},t) - \lambda \, n^2({\bf x},t)
+ \zeta({\bf x},t) \ ,
\end{equation}
where the constant $\tau$ controls the balance between offspring
production
and self-destruction, and plays the role of the deviation $p-p_c$ 
from the critical percolation probability. 
The infection of nearest neighbors is represented
by the diffusion operator $\nabla^2$, while the nonlinear term 
incorporates the exclusion principle on the lattice. The fluctuations
are taken into account by adding a multiplicative 
Gaussian noise field $\zeta({\bf x},t)$ 
which is defined by the correlations
\begin{equation}
\label{Noise}
\langle \zeta({\bf x},t) \zeta({\bf x}',t') \rangle =
2 \Gamma \, n({\bf x},t) \, \delta^d({\bf x}-{\bf x}') \delta(t-t') \ .
\end{equation}
In order to generalize this Langevin equation to the case of 
anomalous DP, the short-range diffusion has to be replaced 
by a non-local integral expression which describes 
long-range spreading according to the
probability distribution $P(r)$:
\begin{eqnarray}
\label{IntegralLangevin}
\frac{\partial}{\partial t} n({\bf x},t) 
&=&
\tau \, n({\bf x},t) - \lambda \, n^2({\bf x},t) + \zeta({\bf x},t)
\\& &
+ D \int d^dx' \, P(|{\bf x}-{\bf x'}|) [n({\bf x'},t)-n({\bf x},t)] \ .
\nonumber
\end{eqnarray}
The two contributions in the integrand describe gain and
loss processes, respectively.
Keeping the most relevant terms in a small momentum expansion,
this equation may be written as~\cite{janssn}
\begin{eqnarray}
\label{Langevin}
\frac{\partial}{\partial t} n({\bf x},t) &=&
\Bigl(D_N \nabla^2 + D_A \nabla^\sigma + \tau \Bigr) n({\bf x},t) 
\\ && \nonumber
- \lambda n^2({\bf x},t) + \zeta({\bf x},t) \ ,
\end{eqnarray}
where the noise correlations are assumed to be the same
as in Eq.~(\ref{Noise}). $D_N$ and $D_A$ are the rates for 
normal and anomalous diffusion, respectively. The anomalous diffusion 
operator $\nabla^\sigma$ describes moves over long distances 
and is defined through its action in momentum space
\begin{equation}
\label{AnomalousDiffusion}
\nabla^\sigma \, e^{i {\bf k} \cdot {\bf x}} =
- k^\sigma \, e^{i {\bf k} \cdot {\bf x}} \ ,
\end{equation}
where $k=|{\bf k}|$. 
The standard diffusive term $D_N \nabla^2$ takes into account the
short range component of the L\'evy distribution. Note that even if this 
term were not initially included, it would still be generated under 
renormalization of the theory.

Before summarizing the field-theoretic results, 
let us first consider the mean-field approximation.
As in ordinary DP the mean-field 
dynamic phase transition occurs at 
$\tau=0$, where gain and loss processes balance one another.
For $\tau<0$, the particle density decays exponentially quickly towards
$n=0$, which is the absorbing state of the system. However, for
$\tau>0$, the stable stationary state now has the non-zero particle
density $n=\tau/\lambda$. Since $\tau$ plays the role
of $p-p_c$, the mean field density
exponent is $\beta^{MF}=1$. The scaling exponents
$\nu_\perp$ and $\nu_{||}$ can be derived from
an inspection of Eq.~(\ref{Langevin}).
For $\sigma<2$, we see that
\begin{equation}
\xi_{\perp}\sim|\tau|^{-\nu_{\perp}} \ , \qquad
\nu^{MF}_{\perp}=1/\sigma \ ,
\end{equation}
and the characteristic time diverges according to
\begin{equation}
\xi_{\parallel}\sim|\tau|^{-\nu_{\parallel}} \ , \qquad 
\nu^{MF}_{\parallel}=1 \ .
\end{equation}
As expected, for $\sigma\geq 2$, these exponents 
cross over smoothly to the ordinary DP exponents.
Note that the mean field result demonstrates that
$\nu_\perp$ varies continuously with $\sigma$. 

The mean field approximation is expected to be quantitatively
accurate above the upper critical dimension. For $d\leq d_c$, however, 
fluctuation effects have to be taken into account.
The fluctuation corrections to the critical exponents can
be computed by a field-theoretic RG calculation. 
Using standard techniques, the Langevin 
equation~(\ref{Langevin}) 
can be rewritten as an effective action:
\begin{eqnarray}
\label{EffectiveAction}
S[\bar{\psi},\psi]
&=& 
\int d^dx ~ dt ~
\Bigl[\bar{\psi}(\partial_t - \tau - D_N \nabla^2 - D_A
\nabla^\sigma)\psi 
\nonumber
\\ && \hspace{20mm}
+  \frac{g}{2}(\bar\psi\psi^2-\bar{\psi}^2\psi) \Bigr] \ .
\end{eqnarray}
Simple power counting on this action reveals that the upper critical
dimension is $d_c=2\sigma$, below which 
fluctuation effects become important.
In the above action the field $\psi({\bf x},t)$ 
can be identified with the
coarse-grained particle density field $n({\bf x},t)$ \cite{howard} and
$\bar{\psi}({\bf x},t)$ is the corresponding response field.
The expression in Eq.~(\ref{EffectiveAction}) differs
from the usual action of Reggeon field
theory~\cite{regfth1,regfth2,regfth3}
by the addition of a term representing anomalous diffusion.

The field-theoretic RG calculation in Ref.~\cite{janssn} employs 
Wilson's momentum shell renormalization group recursion 
relations in order to determine the critical exponents. 
The authors of the present work have independently
performed similar calculations based on dimensional regularization
which are fully consistent with Ref.~\cite{janssn}.
In the following we summarize the main results.
The critical exponents to one-loop order in 
$d=2\sigma-\epsilon$ dimensions 
are given by
\begin{eqnarray}
\label{CriticalExponents}
\beta &=& 1-\frac{2\epsilon}{7\sigma} + O\left( \epsilon^{2}\right) \ ,
\nonumber \\
\nu_{\perp} &=& \frac{1}{\sigma} + \frac{2\epsilon}{7\sigma^2}
 + O\left( \epsilon^{2}\right) \ ,
\nonumber \\
\nu_{||} &=& 1 + \frac{\epsilon}{7\sigma} + O\left( \epsilon^{2}\right)
\ ,
\\
z={\nu_{\parallel}\over\nu_{\perp}}
&=&\sigma-\epsilon/7  + O\left( \epsilon^{2}\right)\nonumber \ .
\end{eqnarray}
Moreover, it can be shown that the hyperscaling relation 
\begin{equation}
\label{HyperScalingRelation}
\theta + 2 \delta = d/z 
\, \qquad
(\delta = \beta/\nu_{||}) \ ,
\end{equation}
for the so-called critical initial 
slip exponent $\theta$~\cite{crslip},
holds for arbitrary values of $\sigma$. 
The exponent $\theta$ (which is also sometimes denoted by 
$\eta$ in the literature)
describes the initial increase in the number of active particles
$N(t)$ for critical systems starting from initial states at
very low density, i.e. where we have $N(t)\sim t^{\theta}$. 
The critical initial slip plays an important role 
in dynamical Monte-Carlo simulations (see Section~\ref{Numerics}). 
To one-loop order, $\theta$ and $\delta$ are given by
\begin{equation}
\label{EtaDelta}
\theta=\frac{\epsilon}{7\sigma} + O\left( \epsilon^{2}\right)
\ , \qquad
\delta=1-\frac{3\epsilon}{7\sigma} + O\left( \epsilon^{2}\right)
\ .
\end{equation}
Finally, thanks to the fact that $D_A$ does not get renormalized,
one can prove the {\em exact} scaling relation
\begin{equation}
\label{NewScalingRelation}
\nu_{||} - \nu_\perp(\sigma-d)-2\beta = 0 \ .
\end{equation}
The fact that $D_A$ does not get renormalized means that anomalous DP is 
described by {\em two} rather than
three independent critical exponents. 
The scaling relation~(\ref{NewScalingRelation}) has a further
surprising consequence. Assuming that $\beta$, $\nu_\perp$
and $\nu_{||}$ change continuously with $\sigma$, then for fixed $d$, 
it predicts the value $\sigma_c$ where the system should cross
over to ordinary DP (assuming the crossover is smooth). 
To this end one simply has to insert
the numerically known values of the DP exponents into 
Eq.~(\ref{NewScalingRelation}). Surprisingly one obtains
$\sigma_c=2.0766(2)$ in one, $\sigma_c\simeq 2.2$ in two,
and $\sigma_c=2+\tilde\epsilon/12$ in $d=4-\tilde\epsilon$ spatial 
dimensions. Thus the crossover takes place at $\sigma_c>2$ which 
collides with the intuitive argument that the anomalous
diffusion operator $\nabla^\sigma$ should only be relevant if
$\sigma<2$. But, as pointed out in Ref.~\cite{janssn}, this
naive argument may be wrong in an interacting theory where the
critical behaviour is determined by a nontrivial fixed point of
an RG transformation. Rather the field-theoretic calculation
predicts that anomalous diffusion is still relevant in the
range $2 \leq \sigma < \sigma_c(d)$ for $d<4$. This prediction seems 
to be additionally surprising since the operators 
for anomalous and ordinary diffusion 
$\nabla^\sigma$ and $\nabla^2$ are 
expected to coincide for $\sigma=2$. 
However, one can show that for 
$\sigma=2$ the most relevant terms in a small 
momentum  expansion of Eq.~(\ref{IntegralLangevin}) also contain
a logarithmic  correction of the form $-k^2 \log k$. Therefore
anomalous and ordinary diffusion are indeed different in that case, 
supporting the view that long-range spreading might be relevant in the 
regime $2 \leq \sigma  < \sigma_c$. Unfortunately, the numerical 
simulations presented in Section~\ref{Numerics} are not accurate 
enough to confirm this prediction.

Another interesting aspect of anomalous DP is that
$\sigma$ can be chosen in such a way that the critical dimension
$d_c=2\sigma$ approaches the actual physical dimension at which 
the simulations are performed (see Section~\ref{Numerics}).
Even in one spatial dimension this allows us to
verify the one-loop results~(\ref{CriticalExponents}). 
For example, if $\sigma=1/2+\mu$, the critical 
dimension of the system is $d_c=1+2\mu$ and hence 
the exponents in a 1+1-dimensional system 
change to first order in $\mu$ as
\begin{eqnarray}
\label{TheoryPredictions}
\beta &=& 1-8 \mu / 7 + O\left( \mu^{2}\right) \ , 
\nonumber \\
\nu_\perp &=& 2-12 \mu / 7+ O\left( \mu^{2}\right) \ ,
\nonumber \\
\nu_{||} &=& 1+4 \mu / 7+ O\left( \mu^{2}\right) \ ,
\\
z &=& 1/2 + 5 \mu / 7+ O\left( \mu^{2}\right) \ ,
\nonumber \\
\delta &=& 1-12 \mu / 7+ O\left( \mu^{2}\right) \ ,
\nonumber \\
\theta &=& 4 \mu / 7+ O\left( \mu^{2}\right) \ .
\nonumber
\end{eqnarray}
In Section~\ref{Numerics} we shall demonstrate that this initial
change of the exponents is indeed observed in numerical
simulations.

\section{A lattice model for anomalous directed percolation}
\label{LatticeModel}

\begin{figure}
\epsfxsize=75mm
\centerline{\epsffile{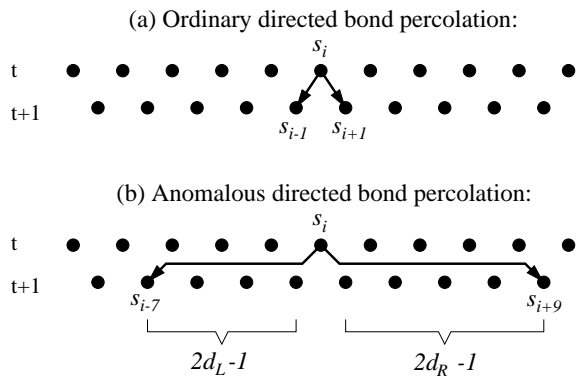}}
\vspace{2mm}
\caption{
\label{FigRules}
Dynamical rules for (a) ordinary directed bond percolation
and (b) the present model with algebraically distributed distances.
$d_L$ and $d_R$ are defined in the text.
}
\end{figure}

Anomalous DP was first studied numerically by Albano~\cite{albano}
who introduced a model for branching-annihi\-lating random walks
in which the particles performed L\'evy flights. However, his estimates 
for the critical exponents were rather inconclusive, in particular 
they violated the scaling relation~(\ref{NewScalingRelation}) 
and even the mean-field limit was not correctly reproduced. 
In Ref.~\cite{janssn} it was suspected that these problems could
have originated in the truncation of the flight distances at some upper 
cutoff, usually at the system size. The upper cutoff effectively
suppressed long range motion and hence DP-like
behaviour was amplified. A systematic finite size analysis of
a similar model confirms this point of view and
shows that even on a lattice with $10^4$ sites finite-size 
effects are still extremely dominant.

Similar problems were also encountered in a more recent study of a
generalized Domany-Kinzel model with long range 
interactions~\cite{cannas}. In this case an upper
cutoff for $P(r)$ was also introduced 
(by defining transition probabilities
$w(S_i^t|S^{t-1})$ in which the sum over the spreading 
distance for a system with $N$ sites is truncated at $N/2$). 
It was reported that the percolation threshold 
depended on the system size and varied by more than
$20$\%. However, it seems that this unusual drift 
of $p_c$ is actually related to extremely strong 
finite size effects. 

In order to minimize finite size effects,
we introduce a model in which the probability 
distribution for long-range spreading is not truncated from above.
As in the case of ordinary directed bond percolation, 
our model is defined on a tilted square 
lattice and evolves by parallel updates. 
A binary variable $s_i(t)$ is attached to 
each lattice site $i$. $s_i=1$ means that the site
is active (infected) whereas 
$s_i=0$ denotes an inactive (healthy) site. 
Although the model may be defined in arbitrary 
spatial dimensions, we will focus here on
the 1+1-dimensional case.
The dynamical rules (see Fig.~\ref{FigRules}) 
depend on two parameters, 
namely the control exponent $\sigma>0$ and the bond 
probability $0\leq p \leq 1$. For a given configuration 
$\{s_i(t)\}$ at time $t$, the next configuration $\{s_i(t+1)\}$ 
is constructed as follows. First the new configuration 
is initialized by setting $s_i(t+1):=0$.
Then a loop over all active sites $i$ in the 
previous configuration is executed. This loop consists of the
following steps:

\begin{figure}
\epsfxsize=85mm
\centerline{\epsffile{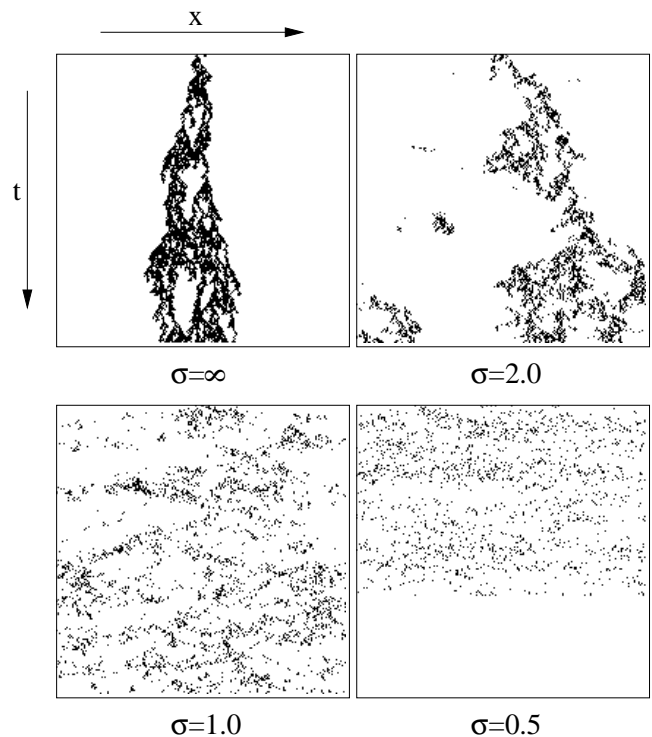}}
\vspace{2mm}
\caption{
\label{FigDemo}
Critical anomalous directed percolation 
in 1+1 dimensions for different 
values of $\sigma$. The figure shows 
typical clusters starting from $5$ active
sites in the center of the lattice.
The case $\sigma=\infty$ corresponds
to ordinary DP. As $\sigma$ decreases, spatial 
structures become more and more smeared out until in
the mean field limit $\sigma=1/2$ the
particles appear to be randomly 
distributed over the whole system. 
For small values of $\sigma$ finite size effects 
may lead to sudden transitions 
into the absorbing state.
}
\end{figure}
\begin{enumerate}

\item   Generate two random numbers $z_L$ and $z_R$ from a flat
        distribution between $0$ and $1$.

\item   Define two real-valued spreading distances 
        $r_L = z_L^{-1/\sigma}$  and $r_R = z_R^{-1/\sigma}$, 
        for spreading to the left~(L) and to the right~(R). The
        corresponding integer spreading distances $d_L$ and $d_R$ 
        are defined as the largest integer numbers that are
        smaller than $r_L$ and $r_R$, respectively. If $d_L$ or $d_R$
exceed
        the allowed range for integer numbers we go back to step~1.

\item   Generate two further random numbers $y_L$ and $y_R$ 
        drawn from a flat distribution between $0$ and $1$, and assign
        $s_{i+1-2d_L}(t+1):=1$ if $y_L<p$, and
        $s_{i-1+2d_R}(t+1):=1$ if $y_R<p$, respectively.
        In finite systems the arithmetic operations in the indices
        are carried out modulo $L$ by assuming periodic boundary 
        conditions, i.e. $s_i \equiv s_{i\pm L}$.

\end{enumerate}

\noindent
The model includes two special cases.
For $\sigma \rightarrow \infty$ it reduces to
ordinary directed bond percolation with $p_c \simeq 0.6447$.
On the other hand, for $\sigma \rightarrow 0$ the interaction
becomes totally random. In that case the model is exactly
solvable and the transition takes place at $p_c=1/2$.
In between, the spreading properties of the model
change drastically, as illustrated in Fig.~\ref{FigDemo}. 

As can be easily verified, the assignment $r = z^{-1/\sigma}$
reproduces the normalized probability distribution
\begin{equation}
P(r) = 
\left\{
\begin{array}{cl}
\sigma/r^{1+\sigma} & \mbox{if \ \ } r>1 \ , \\
0 & \mbox{otherwise} \ .
\end{array}
\right.
\end{equation}
As usual the distribution has a lower cutoff 
at $r_{min}=1$, which represents the lattice spacing. 
But in contrast to previously studied models, 
no upper cutoff is introduced and therefore almost
arbitrarily large spreading distances may be generated 
(limited only by the maximal range of $64$-bit integer numbers). 
In finite systems the target site is determined by
assuming periodic boundary conditions, i.e., the 
particle may ``revolve'' several times around the system.
It turns out that this simple trick considerably reduces 
finite size effects. In particular the value of $p_c$
is well defined over a wide range of system sizes.
Nevertheless finite size effects are  still important 
in this model. 
In particular for small values of $\sigma$, where long-distance
flights occur frequently, finite-size effects enhance the 
probability for a target site to be already occupied. This in turn
reduces the average density of active sites in a growing cluster
and therefore increases the probability to enter the absorbing state.
For example, for $\sigma=0.5$, a small system with
only $200$ sites reaches the absorbing state 
typically after only a few hundred time steps 
(see Fig.~\ref{FigDemo}). Therefore, in
order to further reduce finite size effects, we either 
use very large lattice sizes of about $10^5$ sites 
(as in stationary simulations, see below)
or else in other (dynamical) simulations, we can eliminate
finite size effects almost completely 
by working on a virtually infinite
lattice (see also below).

\section{Numerical results}
\label{Numerics}

\begin{figure}
\epsfxsize=75mm
\centerline{\epsffile{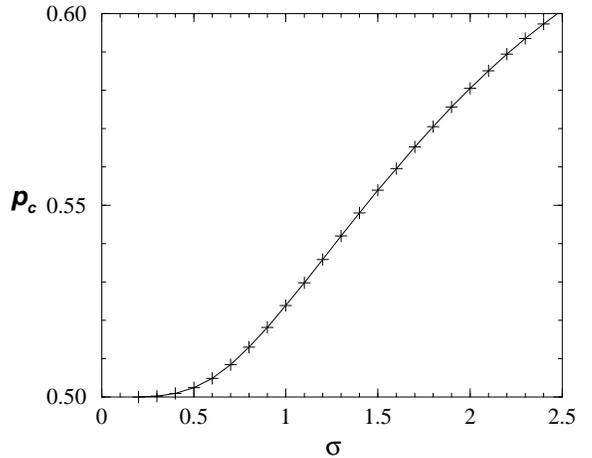}}
\vspace{2mm}
\caption{
\label{FigPc}
The critical percolation threshold of anomalous directed
bond percolation as a function of $\sigma$.
}
\end{figure}
In order to estimate the critical exponents of anomalous DP
we employ two different standard Monte-Carlo techniques,
namely dynamical simulations at criticality 
and steady-state simulations in the active phase.

In {\it dynamical simulations}~\cite{dynsim} a critical cluster 
is grown from a single active seed (just as in Fig.~\ref{FigDemo}). 
Averaging over many independent realizations one measures the 
survival probability $P(t)$, the number of active particles $N(t)$,
and the mean square spreading of surviving clusters from the origin
$R^2(t)$. At criticality, these quantities are expected to scale as 
\begin{equation}
\label{DynamicalScaling}
P(t) \sim t^{-\delta} \ ,
\qquad
N(t) \sim t^\theta \ , 
\qquad
R^2(t) \sim t^{2/z} \ ,
\end{equation}
where $\delta=\beta/\nu_{||}$ and $\theta$ is the critical initial 
slip exponent~\cite{crslip} (see Eq. (\ref{HyperScalingRelation})).
Since the size of the growing cluster is finite, we are able to 
perform the simulations on a virtually infinite lattice by storing 
the coordinates of active particles in a dynamically generated list. 
The effective system size is then determined by the maximal spreading 
range (i.e., the maximal range of integer numbers $\pm 2^{63}$),
which means that finite size effects are almost eliminated.
Since deviations from criticality lead to a curvature of $P(t)$
in a double logarithmic plot, the dynamical simulation method 
allows a precise estimate of the percolation 
threshold $p_c$ for different values of $\sigma$ 
(see Fig.~\ref{FigPc} and Table~\ref{table}). As expected, 
$p_c$ tends to $1/2$ in the limit $\sigma \rightarrow 0$.

\begin{figure}
\epsfxsize=85mm
\centerline{\epsffile{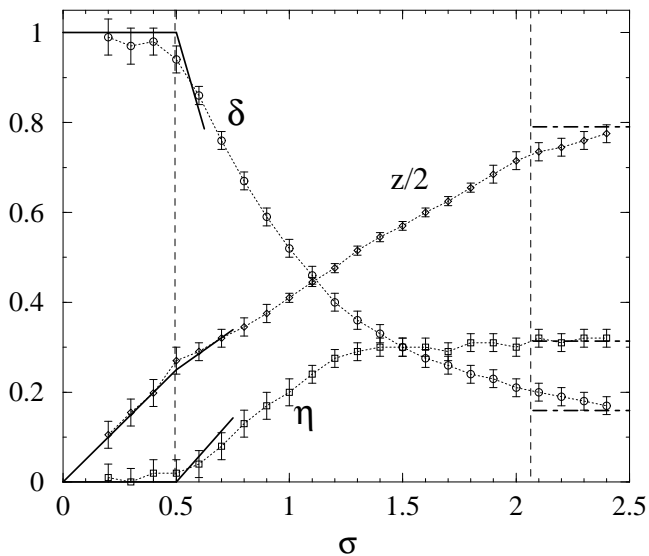}}
\vspace{2mm}
\caption{
\label{FigDynamic}
Estimates for the critical exponents from
dynamical Monte-Carlo simulations in
comparison with the field-theoretic
predictions (solid lines) and the DP exponents
(dot-dashed lines).
}
\end{figure}

Having determined the critical points, 
we measure the quantities $P(t)$, $N(t)$, and $R^2(t)$ 
at criticality. However, it turns out that,
in the presence of sufficiently long-range interactions,
the mean square spreading, defined as an 
{\em arithmetic} average $R^2(t)=\langle |{\bf x}(t)|^2\rangle$, 
diverges. In order to circumvent this problem, we instead compute
the {\em geometric} average
\begin{equation}
R^2(t) = \exp \left[\langle \log(|{\bf x}(t)|^2) \rangle \right] \ .
\end{equation}
This average turns out to be finite for all $\sigma>0$
and renders consistent results in the case of ordinary DP.
The numerical estimates for the dynamical 
exponents $\delta$, $\theta$, and $z/2$ are shown in
Fig.~\ref{FigDynamic}.

\begin{figure}
\epsfxsize=85mm
\centerline{\epsffile{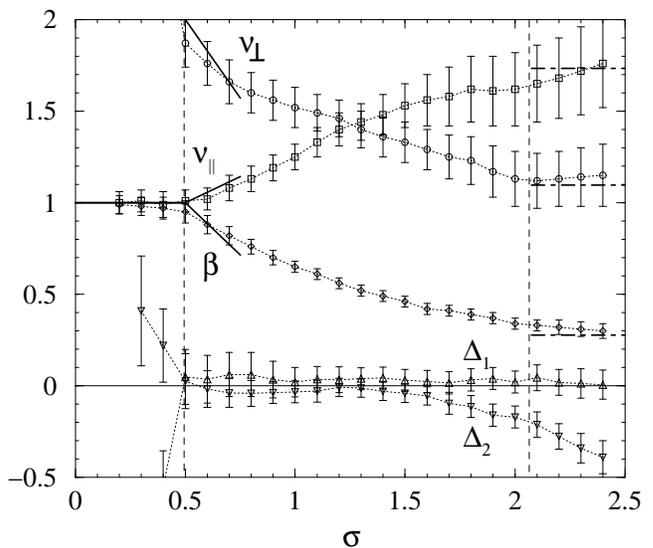}}
\vspace{2mm}
\caption{
\label{FigResults}
Estimates for the exponent $\beta$ and the derived 
exponents $\nu_\perp$ and $\nu_{||}$ in comparison
with the field-theoretic results (solid lines)
and the DP exponents (dot-dashed lines). 
The quantities $\Delta_1$ and
$\Delta_2$ represent deviations from the
scaling relations (\ref{HyperScalingRelation}) and 
(\ref{NewScalingRelation}), respectively (see text).
}
\end{figure}

The exponent $\beta$ is determined by 
{\it stationary simulations} in the active phase.
As the active phase of anomalous DP is characterized 
by a homogeneous particle density, this type of 
simulation has to be performed on a finite lattice.
In order to minimize finite size effects, 
we choose a large lattice size of $L=10^5$ sites.
Starting from a fully occupied initial state,
the system first equilibrates over $10^4$ time
steps before the stationary density $n$ is averaged over
another $10^4$ time steps. Our estimates for $\beta$
are shown in Fig.~\ref{FigResults}. 
Combining the results we can now compute 
the scaling exponents $\nu_\perp=\beta/\delta z$
and $\nu_{||}=\beta/\delta$, which are 
summarized in Table~\ref{table}. 

According to Eq.~(\ref{TheoryPredictions}), the one-loop expansion
predicts the initial variation of the critical
exponents close to $\sigma=1/2$. This is one of the rare cases
where one can directly ``see'' the field-theoretic results 
in the simulation data. In Figs.~\ref{FigDynamic}-\ref{FigResults},
the predicted initial slopes are indicated by solid 
lines. Clearly they are in fair agreement with the numerical 
estimates, which confirms the field-theoretic results of 
Ref.~\cite{janssn}. For $\sigma>1.5$, however, the numerical
results are not accurate enough to verify the predicted 
location of the crossover to ordinary DP at $\sigma_c=2.0766(2)$. 
It seems that the deviations in this regime are due to
very long crossover times in the dynamical simulations,
resulting from a complicated interplay between long-range
and short-range processes.

In order to verify the scaling relations
(\ref{HyperScalingRelation}) and (\ref{NewScalingRelation})
we have also plotted the deviations 
$\Delta_1=2\delta+\theta-1/z$ and
$\Delta_2=1-\sigma+(1-2\delta)z$ which should
be equal to zero in the intervals $\sigma\geq 0.5$ and
$0.5 \leq \sigma \leq \sigma_c$, respectively.
In fact, as shown in Fig.~\ref{FigResults}, $\Delta_1$ is
smaller than the error tolerance, 
which confirms the validity of the hyperscaling 
relation~(\ref{HyperScalingRelation}). 
Similarly the values of $\Delta_2$ confirm 
the validity of Eq.~(\ref{NewScalingRelation})
in the range $0.5 \leq \sigma \leq 1.5$, whereas 
significant deviations occur for $\sigma > 1.5$. We believe
that these deviations do not indicate that the
scaling relation~(\ref{NewScalingRelation}) is violated
for large values of~$\sigma$, rather they 
confirm that the simulations in this regime
may be affected by very long crossover times.

\begin{table}
\begin{tabular}{|c||c|c|c|c|c|c|}
$\sigma$ & $p_c$ & $\beta$ & $\nu_\perp$ & $\nu_{||}$ & $z$ & $\theta$
\\\hline
0.2 & 0.50026(1)  & 0.99(4) & 4.7(6)   & 1.00(6)  & 0.21(2) & 0.01(3)\\
0.3 & 0.50097(1)  & 0.98(5) & 3.2(3)   & 1.01(6)  & 0.31(2) & 0.02(3)\\
0.4 & 0.50245(2)  & 0.97(6) & 2.5(2)   & 0.99(7)  & 0.39(2) & 0.01(3)\\
0.5 & 0.50490(2)  & 0.95(6) & 1.87(13) & 1.01(6)  & 0.54(2) & 0.02(3)\\
0.6 & 0.50847(2)  & 0.88(5) & 1.76(12) & 1.02(6)  & 0.58(2) & 0.04(3)\\
0.8 & 0.51820(3)  & 0.76(4) & 1.60(11) & 1.13(7)  & 0.71(2) & 0.13(3)\\
1.0 & 0.52981(5)  & 0.65(3) & 1.52(11) & 1.25(7)  & 0.82(2) & 0.20(3)\\
1.2 & 0.54197(5)  & 0.56(3) & 1.46(10) & 1.40(9)  & 0.96(3) & 0.28(2)\\
1.4 & 0.55390(10) & 0.49(3) & 1.36(11) & 1.48(11) & 1.09(3) & 0.30(2)\\
1.6 & 0.56520(10) & 0.43(3) & 1.29(11) & 1.56(14) & 1.21(3) & 0.30(2)\\
1.8 & 0.57561(10) & 0.39(3) & 1.23(13) & 1.62(18) & 1.32(3) & 0.31(2)\\
2.0 & 0.58505(10) & 0.34(3) & 1.13(15) & 1.62(20) & 1.43(3) & 0.32(2)\\
2.2 & 0.59345(10) & 0.32(3) & 1.13(15) & 1.68(21) & 1.49(4) & 0.31(2)\\
2.4 & 0.60085(10) & 0.30(3) & 1.15(17) & 1.76(24) & 1.53(5) & 0.32(2)\\ 
\hline
DP  & 0.644700    & 0.2765  & 1.097  & 1.734  & 1.581 & 0.3137 
\end{tabular}
\vspace{1mm}
\caption{
\label{table}
Estimates of the percolation threshold and the critical
exponents for various values of $\sigma$, compared to
the corresponding values for ordinary bond DP.}
\end{table}

\section{Anomalous annihilation process}
\label{Annihilation}
In this section we consider the somewhat simpler case of
anomalous pair annihilation $A+A \rightarrow \emptyset$ with long-range 
hopping. This model was previously studied in \cite{zum}, using both 
simulations and approximate theoretical techniques. In this paper we
will extend this previous work, by presenting a systematic
field-theoretic 
analysis, as well as by performing more detailed numerical simulations. 

In the ordinary annihilation process~\cite{benlee} 
with short-range interactions, the average particle density 
is known to decay as
\begin{equation}
\label{annden}
n(t) \sim \left\{ \begin{array}{ll}
    t^{-d/2} &{\rm for \ } d < 2 \ , \\
    t^{-1} \ln t &{\rm for \ } d = d_c = 2 \ , \\
    t^{-1} &{\rm for \ } d > 2 \ .
    \end{array} \right.
\end{equation}
Hence, except for the $\log$ correction in $d=2$, the density decays 
away as a power law, $n(t)\sim t^{-\alpha}$. 
Turning now to the L\'evy-flight case, this may be described
theoretically
by inserting an additional operator $\nabla^\sigma$ into the well-known
field-theoretic action for pair annihilation (see \cite{benlee}). 
The resulting action reads
\begin{eqnarray}
\label{AnnihilationAction}
S[\bar{\psi},\psi]
&=& 
\int d^dx \, dt \,\Bigl\{
\bar{\psi} (
\partial_t - D_N \nabla^2 - D_A \nabla^\sigma 
)\psi
\nonumber
\\ && \hspace{12mm}
+  2 \lambda \bar{\psi}\psi^2 + \lambda \bar{\psi}^2\psi^2 -
n_0\bar{\psi}\delta(t)   
\Bigr\} \ ,
\end{eqnarray}
where $n_0$ is the initial (homogeneous) 
density at $t=0$. Here the field 
$\psi$ is {\it not} simply related to the 
coarse-grained density field \cite{howard}, 
although it is true that the average 
values of both fields are the same. 
The action (\ref{AnnihilationAction}) 
can be {\it derived} systematically, starting with an 
appropriate (non-local) Master equation --- the details are given in 
Appendix~A. Note also that the action for the process $A+A\to A$ with
L\'evy flight hops differs only in the 
coefficients of the reaction terms.
Hence the L\'evy flight annihilation and 
coagulation processes are in the
same universality class.

An analysis of the above action follows very closely that
of Ref.~\cite{benlee}. For $\sigma<2$,
power counting reveals that the upper critical dimension
of the model is now $d_c=\sigma<2$. 
For $d>d_c$ mean-field theory is expected
to be quantitatively accurate, with an 
asymptotic density decay $\sim t^{-1}$.
Below the upper critical dimension, however, 
the renormalized reaction rate 
flows to an order $\epsilon=\sigma-d$ 
fixed point. This allows us to very 
quickly determine the asymptotic density 
decay via dimensional arguments. 
Below $d_c$ the only dimensionful quantity 
left in the problem is the time $t$,
which, for $\sigma<2$, scales as $[t]\sim k^{-\sigma}$. 
Hence, for $\sigma<2$, the density must decay as:
\begin{equation}
\label{AnnhilationResult}
n(t) \sim \left\{ \begin{array}{ll}
    t^{-d/\sigma}       & {\rm for \ } d<\sigma \ , \\
    t^{-1} \ln t             & {\rm for \ } d=d_c=\sigma \ , \\
    t^{-1}              & {\rm for \ } d>\sigma \ .
    \end{array} \right.
\end{equation}
The derivation of the logarithm at the upper critical dimension 
requires a slightly more sophisticated calculation, which is, however, 
completely analogous to that in Ref.~\cite{benlee}. 
Note also that for $\sigma\geq 2$ the results cross over smoothly
to the standard annihilation exponents of Eq.~(\ref{annden}). 

A lattice model for anomalous annihilation 
in $1+1$ dimensions may be constructed by a simple modification 
of the model for anomalous DP introduced in Section~\ref{LatticeModel}. 
To this end steps $1$-$3$ have been modified such that for all active 
sites $i$ we perform the following procedure:
\begin{enumerate}
\item   Generate a random number $z \in [0,1]$ and
        define a real-valued spreading distance 
        $r = z^{-1/\sigma}$. The corresponding 
        integer spreading distance $d$ is defined 
        as the largest integer number smaller than $r$.

\item   Generate another random number $y \in [0,1]$ and
        assign $s^{\rm new}_{i+1-2d}(t+1):=1-s^{\rm old}_{i+1-2d}(t+1)$ 
        if $y>1/2$, and $s^{\rm new}_{i-1+2d}(t+1):=1-s^{\rm
old}_{i-1+2d}
        (t+1)$ otherwise.
        As in the case of anomalous DP, 
        the arithmetic operations in the indices
        are carried out modulo $L$ by assuming periodic boundary 
        conditions, i.e. $s_i \equiv s_{i\pm L}$.
\end{enumerate}
In step 2 the state of the target site is simply inverted. Therefore,
if two particles move to the same target site, they annihilate
instantaneously.

\begin{figure}
\epsfxsize=85mm
\centerline{\epsffile{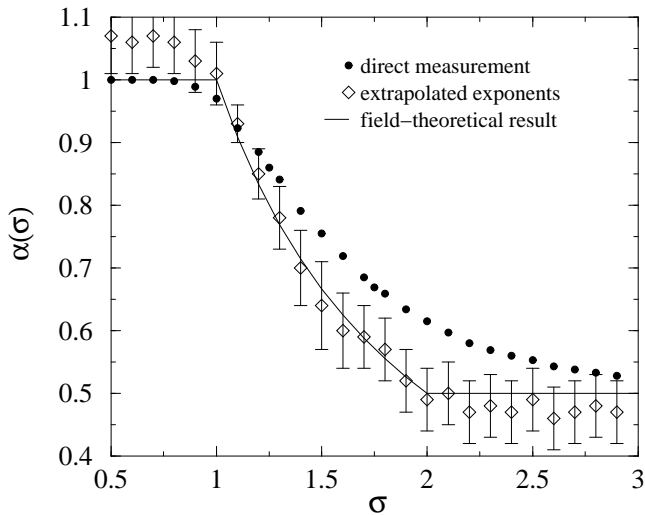}}
\vspace{2mm}
\caption{
\label{FigAnnh}
The anomalous annihilation process: the graph shows direct
estimates and extrapolations for the decay exponent $\alpha$,
as a function of $\sigma$. The solid line represents
the exact result (neglecting $\log$ corrections at $\sigma=1$).}
\end{figure}

Since the annihilation process starts with 
a homogeneous density of particles,
it is impracticable to work with a virtually 
infinite lattice by storing the
coordinates of individual particles in a list. 
Rather we have to perform
ordinary simulations with a fixed system size. 
In order to minimize finite
size effects we choose a large lattice size of $2^{16}$ sites.
For various values of~$\sigma$ we measure the particle density $n(t)$
up to $10^4$ time steps averaged over at least $10^3$ independent runs.
By measuring the slopes of $n(t)$ in a double logarithmic plot in the
decade $10^3 \leq t \leq 10^4$, we estimate the density decay exponents 
$\alpha(\sigma)$,
which are shown in Fig.~\ref{FigAnnh} (labeled as direct measurements).
For $\sigma<1.5$ the agreement with the theoretical result 
of Eq.~(\ref{AnnhilationResult}) (the solid line) is quite convincing,
whereas large deviations occur close to $\sigma=2$. 
A detailed analysis of the local slope of $n(t)$ as a function of time
in a double-logarithmic plot shows that these deviations are related 
to very long crossover times. In fact, determining the local 
slopes of $n(t)$ in a log-log plot and extrapolating 
them graphically to $t \rightarrow \infty$, 
one obtains a much better coincidence. On the other hand, 
for $\sigma<1$, the extrapolation leads to larger deviations.
These errors may be related to finite size effects which are still
extremely dominant in this regime.
%
%

\section{Conclusions}
\label{Conclusions}
In this paper we have, for the first time, 
studied numerically the behaviour
of L\'evy-flight DP close to the phase transition between the active and
absorbing states. To this end, we have introduced a lattice model for
anomalous DP in which finite size effects are considerably reduced. 
Using special simulation techniques we have obtained accurate values for 
the associated critical exponents in $1+1$ dimensions, which are almost 
free from finite-size effects. 
In addition we have performed quantitative 
tests of the one loop field-theoretic results, 
by tuning the upper critical
dimension of the model to lie just above $d=1$. 
Our results are all in good
agreement with the recent field-theoretic analysis of~\cite{janssn}.
Close to $\sigma=2$, however, our numerical results are not accurate 
enough to confirm the form of the predicted crossover to ordinary DP.
We have also considered the simpler case of L\'evy-flight pair
annihilation,
where our numerics are again in agreement 
with (exact) field-theoretic  arguments.

Various possible extensions of the above 
models are possible. The most obvious
involves including a power law waiting 
time distribution for the particles,
in addition to the power law L\'evy 
distribution for particle hops. Hence,
in the absence of interactions, 
each particle would perform a {\it continuous
time L\'evy flight} (see \cite{fogedby} and references therein). This
modification should lead to a further universality class, with the
exponents depending continuously on the control parameters for both the 
L\'evy-flight and the (power law) waiting time distributions.


\appendix
\section{Derivation of the anomalous annihilation action}

In this appendix we briefly describe how 
the anomalous annihilation action
(\ref{AnnihilationAction}) may be derived. 
To simplify matters we will not
include the reaction terms --- their derivation 
is precisely the same as in
\cite{benlee}. The appropriate Master 
equation for (anomalous) diffusion is given by
\begin{eqnarray}
\label{master}
& & {\partial\over\partial t}P(\{n\};t)={D_0\over
l^{\sigma}}\sum_i\sum_{j(\neq i)}
\left[(n_j+1)q_{ji} \right. \\
& & \left.
\times P(\ldots,n_j+1,\ldots,n_i-1,\ldots;t)-n_i\, q_{ij}\, P(\{n\};t)
\right] \ , \nonumber
\end{eqnarray}
where $P(\{n\})$ is the probability of 
the particle configuration
$\{n\}=(n_1,\ldots,n_N)$, 
$l$ is the microscopic lattice spacing, 
$D_0$ is the diffusion constant, 
and where $q_{ij}$ gives the appropriate weight for a hop from
site $i$ to site $j$. Following \cite{Peliti}, we next introduce
creation and annihilation operators $a_i$, $a_i^{\dag}$, such that
\begin{equation}
a_i^{\dag}|n_i\rangle=|n_i+1\rangle \ , \qquad a_i|n_i\rangle=
n_i|n_i-1\rangle \ , \\
\end{equation}
with the commutator $[a_i,a_j^{\dag}]=\delta_{ij}$. The system state is 
then given by
\begin{equation}
|\Phi(t)\rangle=\sum_{n_1,\ldots ,n_N}P(\{n\};t)a_1^{\dag^{n_1}}\ldots
a_N
^{\dag^{n_N}}|0\rangle \ ,
\end{equation}
where $|0\rangle$ is the vacuum state. Hence we can rewrite
Eq.~(\ref{master}) as
\begin{equation}
{\partial\over\partial t}|\Phi(t)\rangle=-{\cal H}|\Phi(t)\rangle \ ,
\end{equation}
where
\begin{equation}
{\cal H}=-{D_0\over l^{\sigma}}\sum_i a_i^{\dag}\left[\sum_{j(\neq
i)}(a_j-a_i)
q_{ji}\right] \ .
\end{equation}
We may now perform the mapping to a field theory using standard 
methods (see \cite{Peliti}). After taking the continuum limit in space,
we end up with the continuum action
\begin{eqnarray}
& & S=\int
d^dx~dt~\bigl\{\hat\psi(x,t)~\partial_t\psi(x,t)-D_1~\hat\psi(x,t)
\int d^dy \nonumber \\ 
& & \qquad\qquad\qquad\times([\psi(y,t)-\psi(x,t)] f|x-y|)\bigr\} \ ,
\end{eqnarray}
where we have the L\'evy distribution
\begin{equation}
f(r)~d^dr\sim{1\over r^{\sigma+d}}~d^dr \ .
\end{equation}
Transforming this into Fourier space, we obtain
\begin{eqnarray}
& & 
S=\int {d^dk\over (2\pi)^d} ~ dt ~ \bigl\{
\tilde{\hat\psi}(k,t)\partial_t
{\tilde\psi}(-k,t)-D_1[{\hat{\tilde\psi}}(k,t) 
\nonumber \\
& & \qquad\qquad\qquad\qquad\qquad \times (f(k)-1){\tilde\psi}(-k,t)]
\bigr\} \ ,
\end{eqnarray}
with
\begin{equation}
f(k)-1={1\over{\cal N}}\int_l 
d^dr~{(e^{-ik.r}-1)\over r^{d+\sigma}} \ ,
\end{equation}
where $\cal N$ is a normalization constant.
After some manipulation of the above 
integral, and after performing a
small momentum expansion, we end up with 
\begin{eqnarray}
& & S=\int {d^dk\over (2\pi)^d}~dt~\bigl[{\tilde{\hat\psi}}(k,t)
\partial_t\tilde\psi(-k,t)+ \nonumber \\
& & \qquad +{\tilde{\hat\psi}}(k,t)\{D_A k^{\sigma}+D_N k^2+O(k^4)\}
\tilde\psi(-k,t)\bigr] \ ,
\end{eqnarray}
valid for $0<\sigma<4$, $\sigma\neq 2$. The final action 
(\ref{AnnihilationAction}) is then obtained by the
inclusion of both the reaction terms, and the initial density source.



\begin{thebibliography}{99}

\bibitem{mollis}
                D.~Mollison, J. R. Stat. Soc. B {\bf 39}, 283 (1977).

\bibitem{liggett}
                T.~M.~Liggett, {\it Interacting Particle Systems},
                (Springer, New York, 1985).
                
\bibitem{cataly}
                R.~Ziff, E.~Gulari, and Y.~Barshad,
                Phys. Rev. Lett. {\bf56}, 2553 (1986).

\bibitem{fires}
                E.V.~Albano, J. Phys. A {\bf 27}, L881 (1994).
                
\bibitem{rmedia1}
                S.~Havlin and D.~ben-Avraham, Adv. Phys. {\bf 36}, 695
(1987).

\bibitem{rmedia2}
                J.-P.~Bouchaud and A.~Georges, Phys. Rep. {\bf 195}, 127 
                (1990).
                                
\bibitem{dpreview} 
                W.~Kinzel, in
                        {\em Percolation Structures and Processes},
                        ed. G.~Deutscher, R.~Zallen, and J.~Adler,
                        Ann. Isr. Phys. Soc. {\bf 5}
                        (Adam Hilger, Bristol, 1983), p.~425.

\bibitem{regfth1} 
                P.~Grassberger and K.~Sundermeyer, 
                        Phys. Lett. B {\bf 77}, 220 (1978).
                
\bibitem{regfth2}
                J.L.~Cardy and R.L.~Sugar, 
                        J. Phys. A {\bf 13}, L423 (1980).
\bibitem{regfth3}
                H.K.~Janssen, Z. Phys. B {\bf 42}, 151 (1981).

\bibitem{dpconj} 
                P.~Grassberger, Z. Phys. B {\bf 47}, 365 (1982).
                
\bibitem{grassb}
                P.~Grassberger, in {\it Fractals in physics}, eds.
                L.~Pietronero and E. Tosatti (Elsevier, 1986).

\bibitem{janssn}
                H.K.~Janssen, K.~Oerding, F.~van~Wijland, and H.J.~Hilhorst,
                preprint cond-mat/9807155, 
		submitted to Euro. Phys. J. {\bf B}.

\bibitem{albano}
                E.V.~Albano, Europhys. Lett. {\bf 34}, 97 (1996).
                
\bibitem{cannas}
                S.A.~Cannas,
                preprint cond-mat/9711297.

\bibitem{howard}
                M.J.~Howard and U.C.~T\"auber, J. Phys. A {\bf 30}, 7721
                (1997).

\bibitem{crslip}
                H.K.~Janssen, B.~Schaub, and B.~Schmittmann,
                        Z. Phys. B {\bf 73}, 539 (1989); 
                H.W.~Diehl and U.~Ritschel, 
                        J. Stat. Phys. {\bf 73}, 1 (1993); 
                F.~van~Wijland, K.~Oerding, and H.J.~Hilhorst,
                        Physica A {\bf 251}, 179 (1998).

\bibitem{dynsim}
                P.~Grassberger and A.~de~la~Torre,
                Ann. Phys. (N.Y.) {\bf 122}, 373 (1979).

\bibitem{zum}
                G.~Zumofen and J.~Klafter, 
			Phys. Rev. E {\bf 50}, 5119 (1994).

\bibitem{benlee}
                B.P.~Lee, J. Phys. A {\bf 27}, 2633 (1994).

\bibitem{fogedby}
                H.C.~Fogedby, Phys. Rev. E {\bf 50}, 1657 (1994).

\bibitem{Peliti}
                L.~Peliti, J. Physique {\bf 46}, 1469 (1985).

\end{thebibliography}
\end{document}